\documentstyle[12pt]{article}

\def\pa{\partial}

\def\a{\alpha}

\def\d{\delta}

\def\G{\Gamma}

\def\g{\gamma}

\def\m{\mu}
\def\n{\nu}

\def\mn{{\mu\nu}}

\def\be{\begin{equation}}
\def\bea{\begin{eqnarray}}
\def\ee{\end{equation}}
\def\eea{\end{eqnarray}}

\begin{document}

\hfill BRX TH-549

\begin{center}
{\large\bf Why is the metric invertible?}

{\large S.\ Deser}

Department of Physics, Brandeis University,
Waltham, MA 02454\\and\\
Lauritsen Laboratory, California Institute of Technology,
Pasadena, CA 91125

\end{center}

{\abstract We raise, and provide an (unsatisfactory) answer to,
the title's question: why, unlike all other fields, does the
gravitational ``metric" variable not have zero vacuum? After
formulating, without begging it, we exhibit additions to the
conventional action that express existence of the inverse through
a field equation.}

The metric variable's invertibility is (on a par with its
dimensionality and signature) a tacit, but basic, assumption of
gravitational theories.  Unlike all other fields' dynamical
variables, it does not vanish in the ground (or any other) state.
This property is taken for granted (but see \cite{wilczek} for a
recent attempted explanation)  because it underlies existence of
geometry and because the background-independence of covariant
models does not single out any natural ``zero". Nevertheless, even
if spacetime is but an emergent property of some substrate, one
should still seek an intrinsic explanation of ``why there is
something rather than nothing". Ours will exhibit additions to
gravitational actions that embody invertibility as a field
equation.  The result will be far less satisfactory (or at least
less familiar) than the Higgs effect's construction of a
non-vanishing VEV.

The first difficulty is just to establish a framework where
invertibility is not presupposed; we invoke the Palatini,
first-order, approach where metric and affinity are independent
variables.  For concreteness, consider ordinary GR in D=4,
 \be
 L_E = g^\mn R_\mn (\G ) \; .
 \ee
Here $g^\mn$ is a symmetric contravariant {\it density},
$\G^\a_\mn$ is likewise $(\mn )$ symmetric, and $R_\mn(\G )$ is
the usual affine Ricci tensor constructed from $\G$; note that (at
D=4) $\sqrt{-\det \: g^\mn}$~ is a scalar density.

The Palatini procedure requires solving the field equation
 \be
 D_\a (\G ) g^\mn \sim \pa_\a g^\mn + g \G = 0
 \ee
 for $\G (g)$.  This is where invertibility of $g^\mn$ comes in: without it,
$\G$ remains undetermined.  Second order, Hilbert-Einstein,
actions where $\G$ is already the metric affinity obviously cannot
even be written, absent invertibility.

Having pinpointed the requirement's origin, we provide the most
elementary formal enlargement of the action (1) so that
invertibility becomes a consequence of the field equations.  The
simplest way to ensure that $g^\mn$ is nonsingular is of course
that its determinant not vanish.  [In this connection, we
emphasize that our task is obtaining the inequality $\sqrt{-g}
> 0$, rather than imposing (as in \cite{wilczek}) the condition
$\sqrt{-g} =1$.]  To this end, we append the term
  \be
  L_1 = M \Big(\sqrt{-g} \: \sqrt{-h} - 1\Big)
  \ee
to $L_E$.  We have been forced to introduce a new (covariant
anti-density) tensor field, $h_\mn$, although a simple scalar
anti-density $h$ would have sufficed.  With either choice, one
finds that $\sqrt{-g} \: \sqrt{-h} = 1$, as desired to guarantee
invertibility.  Then varying $h$ implies that $M$=0, since
$\sqrt{-g} \neq 0$.  The basic mechanism of (3) can be made more
elaborate in
 \be
 L_2 = M^\n_\m (g^{\m\a} h_{\a\n} - \d^\m_\n ) \; .
 \ee
 Varying the mixed tensor $M^\m_\n$ ensures that $g^\mn$ has an
 inverse, while the contributions from $g-$ and $h-$variations
 vanish. Once we have a ``normal" metric at hand, we can use
 it to move indices as usual.  We then learn that the symmetric
 part of $M_\mn$ -- the only part that enters in the field
 equations -- vanishes.  A final variant is the pseudo-Higgs
  \be
  L_3 = \tilde{M} \Big( \sqrt{gh} - \phi^2 \Big)
  \ee
  where the Higgs scalar $\phi$ has a non-zero VEV.  This is,
however, a rather empty ``improvement" since one can absorb the
$\phi^2$ into the density $\tilde{M}$ and the metrics.  None of
these ans\"{a}tze provide any physical insight into the origin of
the invertibility, even though they accomplish our formal aim: the
Lagrange multipliers remain proxies for a worthier explanation.

It may be instructive to contrast the above background-free
framework to that of the self-interaction bootstrap of
\cite{deser} in which free tensor gauge field $(h^\mn , \:
\G^\a_\mn )$ propagates on a fixed but arbitrary background
geometry, with an invertible metric $g^\mn_0$ and its associated
connection $\g^\a_\mn (g_0)$. Its action is just the linearization
of (1) about $g^\mn_0$, with $g=g_0 + h$. For consistency of the
resulting field equations, the background is required to be
Ricci-flat: $R_\mn (\g (g_0)) =0$; the $\G (h)$ relation (2)
becomes perfectly soluble,
  \be
  D_\a h^\mn = (g_0\G )^\mn_\a \;\; \Rightarrow \;\; \G = g^{-1}_0 D h \;
  ,
  \ee
 where $D_\a$ is the background covariant derivative.
The full Einstein theory for $g = g_0 + h$ is simply recovered
upon restoring the cubic $L_3 \sim h \G\G$ term,
 modifying (6) to
  \be
 D_\a h^\mn = [(g_0 + h ) \G ]^\mn_\a \; .
 \ee
For generic $h^\mn$, the invertibility of $(g^\mn_0 + h^\mn )$ is
protected by that of $g^\mn_0$.  In our previous language, this
procedure consists in a by-hand separation of the metric variable
into a sum, one term of which is simply declared to define a
fixed, invertible, background geometry: the question is still
begged, if less obviously.

I thank R.\ Jackiw and A.\ Schwimmer for very patient discussions.
This work was supported in part by NSF grant PHY04-01667.

\end{document}